\documentstyle[12pt]{article}

\begin{document}
\def \non{\nonumber}
\def \ra{\rightarrow}
\def \bea{\begin{eqnarray}}
\def \eea{\end{eqnarray}}
\def \Qbar{\overline Q}
\def \qqbar{Q\Qbar}
\def \slj{^{2S+1}L_J}
\def \jpsi{J/\psi}
\def \ppbar{p\overline{p}}
\def \ccbar{c\overline{c}}
\def \pt{$p_T$}
\def \sg{\sigma}
\begin{titlepage}
\phantom{TIFR}
\bigskip
\begin{flushright}\vbox{\begin{tabular}{c}
           TIFR/TH/98-09\\
           August, 1998\\
           hep-ph/9803424\\
\end{tabular}}\end{flushright}
\begin{center}
{\Large \bf $\eta_c$ Production at the Tevatron: A test of NRQCD}
\end{center}
\bigskip
\begin{center}
   {Prakash Mathews\footnote{E-mail: prakash@theory.tifr.res.in},
    P.~Poulose\footnote{E-mail: poulose@theory.tifr.res.in} and
    K.~Sridhar\footnote{E-mail: sridhar@theory.tifr.res.in}\\[1cm]

    {\it
    Department of Theoretical Physics\\
    Tata Institute of Fundamental Research\\
    Homi Bhabha Road, Bombay 400005, INDIA.}}
\end{center}
\bigskip
\begin{abstract}
\noindent We use NRQCD to predict the cross section for $\eta_c$
production at the Tevatron. The non-perturbative matrix elements
required are obtained, using heavy-quark symmetry, from the matrix
elements determined from the CDF $J/\psi$ data. Our numbers are,
therefore, predictions of NRQCD and provide a very good test of it.
Even after taking into account the small branching ratio of the $\eta_c$
into two photons, a substantial rate for the production of this
resonance at the Tevatron is expected.
\end{abstract}
\vspace{1cm}
\begin{center}
\bf (To appear in Physics Letters B)
\end{center}
\end{titlepage}

It is by now well known from compelling theoretical \cite{BBL1} and 
phenomenological \cite{cdf} reasons that the Colour Singlet model 
(CSM) used to describe the production and decay of quarkonium is 
incomplete.  A new factorisation approach within QCD was provided
by Bodwin, Braaten and Lepage \cite{BBL}.  This method  
separates the perturbatively calculable short-distance ($\leq 1/M_Q$, 
$M_Q$ is the mass of the heavy quark) 
effects from the long-distance effects, which are described by 
non-perturbative matrix elements in an effective field theory called 
non-relativistic QCD (NRQCD).  Unlike CSM where factorisation 
is explicitly violated, factorisation can be shown to 
be valid 
in NRQCD, because in this approach the bound state problem is
treated in a systematic expansion in powers of $v$, the relative
velocity between the quarks in the bound state. Consequently, in NRQCD
the production and decay of quarkonium takes place via 
intermediate states in which the $ \qqbar$ pair has quantum numbers 
different from those of the physical quarkonium. The $\qqbar$
pair in the intermediate state denoted as $\qqbar 
[\slj^{[1,8]}]$ could be in the colour singlet or octet 
state and could form the physical quarkonium state by making 
chromo-electric or -magnetic transitions.  
Thus the cross section for production of a quarkonium state $H$ can be 
written as
\bea
\sg(H)=\sum_n\frac{F_n}{M_Q^{d_n-4}}\,\left<0\left|{\cal O}_n^H
	\right|0\right>.
\label{factorizn}
\eea
The coefficients $F_n$ correspond to the production of $\qqbar$ in the 
angular momentum and colour state denoted by $n$ and is calculated using 
perturbative QCD.  The non-perturbative part, $\langle{\cal O}_n^H 
\rangle$ of mass dimension $d_n$ in NRQCD has a well-defined operator 
definition and is universal.  These matrix elements can be extracted from 
any one process and then be used to predict other processes where 
the same matrix elements appear.  Though the summation involves an infinite
number of terms, the relative magnitude of the various terms is
predicted by NRQCD as these matrix elements scale as powers of $v$. 
However, this does not necessarily imply that effects from higher
orders in $v$ will always be small in physical processes, because
any observable, like the decay width or the cross
section is given by a double expansion in the strong coupling constant 
$\alpha_s(M_Q)$ 
and the relative velocity $v$. Indeed, the higher order colour-octet
components have important effects in certain processes: for example,
in $\chi_c$ decays into light hadrons, where these octet components are 
necessary to obtain an infra-red safe expression for the decay-width
\cite{BBL}, and in large-$p_T$ $J/\psi$ and $\psi^\prime$ production from
the CDF experiment \cite{cdf} at Tevatron where discrepancies
larger than one order of magnitude $vis\ a\ vis$ CSM predictions could
be explained by including the octet components \cite{frag,brfl}. 

The non-perturbative matrix elements in NRQCD, are not calculable 
and have to be obtained by fitting to available data. The matrix 
elements of the colour-singlet operators are related to the radial 
wave-functions and can be obtained from decay widths or lattice
calculations, for example, but the colour-octet matrix elements are not
known and have only been obtained by fitting NRQCD predictions to the
CDF data \cite{cho1, cho2}. Given that the normalisation of the cross sections
are not predictions of the theory, it is important to look for other 
independent quantitative tests of NRQCD. One crucial test of NRQCD
was thought to be the production of large-$p_T$ $inelastic$ $J/\psi$ production
in $ep$ collisions at HERA. The inelastic cross section has been
measured by both H1 and ZEUS \cite{h1zeus}. 
The inelasticity of the events is ensured by selecting events with $z<1$,
where $z\equiv p_p\cdot p_{J/\psi}/p_p\cdot p_{\gamma}$. In addition,
a $p_T$ cut of 1~GeV is used to select the events.
Using the values of
the non-perturbative inputs obtained from the Tevatron, the theoretical
predictions have been compared \cite{cakr} to these inelastic data.
However, there seems to be a problem in predicting the $z$ distribution
because while the colour-singlet cross section dominates in most of the
low-$z$ region, the colour-octet contribution grows anomalously in the 
large-$z$ ($0.8<z<0.9$)
region and this rise is not seen in the data. However, it is premature to
conclude that this is a failure of NRQCD, because at $z$ values of 0.8
other effects like soft-gluons \cite{softg}, $k_T$-smearing \cite{smear} could
become important, or even worse, the NRQCD factorisation may break down 
\cite{wise}. The safer conclusion is that the inelastic $J/\psi$ production
process at HERA is not a clean test of NRQCD. 

In fact, NRQCD is best tested at Tevatron itself in other large-$p_T$
observables. The polarisation of the produced $J/\psi$ is one such
test \cite{trans}. The second test is to look for the production of other
charmonium states at large-$p_T$ at the Tevatron. This is the philosophy
of the present paper and we study large-$p_T$ $\eta_c$ production 
at the Tevatron,
with this end in mind. The remarkable thing is that the non-perturbative
parameters appearing in the $\eta_c$ production cross section can be
determined from the matrix elements determined from $J/\psi$ production
at the Tevatron: this happens because of the heavy-quark symmetry of
the NRQCD Lagrangian. This has been exploited earlier in the context
of $h_c$ production at Tevatron \cite{sri}. We show, in the following,
that from NRQCD one expects a large cross section for $\eta_c$ production 
at the Tevatron. The canonical search channel for the $\eta_c$ is its
two photon decay mode and the number of events is substantial even after
taking the small branching ratio of the $\eta_c \rightarrow \gamma \gamma$
into account.

By writing a Fock space expansion of the physical $\eta_c$, which is a $^1S_0$
($J^{PC}=0^{-+}$) state, we have
\bea
\!  \!  \!  \!
\left|\eta_c\right>={\cal O}(1)
	\,\left|\qqbar[^1S_0^{[1]}] \right>+
	 {\cal O}(v^2)\,\left|\qqbar[^1P_1^{[8]}]\,g \right>+
	{\cal O}(v^4)\,\left|\qqbar[^3S_1^{[8]}]\,g \right>+\cdots ~.
\label{fockexpn}
\eea
The colour-singlet $^1S_0$ state contributes at ${\cal O}(1)$ but the
colour-octet $^1P_1$ and $^3S_1$ channels effectively contribute
at the same order because the $P$-state production is itself down
by factor of ${\cal O}(v^2)$.
The colour-octet states become a physical $\eta_c$ by the $^1P_1^{[8]}$ 
state emitting a gluon in an E1 transition, while the $^3S_1^{[8]}$
emitting a gluon in an M1 transition.  To ${\cal O} (\alpha_s^3 v^7)$ the 
$\eta_c$ production cross section is
\bea
\sg(\eta_c)&=&
	\frac{F_1[^1S_0]}{M^2}\, \left< 0 \right| {\cal O}_1^{\eta_c}
		[^1S_0]\left| 0 \right> \nonumber\\
     && + \frac{F_8[^1P_1]}{M^4}\, \left< 0 \right| {\cal O}_8^{\eta_c}
		[^1P_1]\left| 0 \right> 
        + \frac{F_8[^3S_1]}{M^2}\, \left< 0 \right| {\cal O}_8^{\eta_c}
		[^3S_1]\left| 0 \right>.
\eea
The coefficients $F$'s are the cross sections for the production of
$\ccbar$ pair in the respective angular momentum and colour states
and is given by
\bea
\!  \!  \!  \!  \!  \!  \!  \!
\!  \!  \!  \!  \!  \!  \!  \!
\!  \!  \!  \!  \!  \!  \!  \!
\!  \!  \!  \!  \!  \!  \!  \!
&&\frac{d\sg}{dp_T} \;(\ppbar \ra \ccbar\; [^{2S+1}L^{[1,8]}_J]\, X)= \non \\
&&\sum \int \!dy \int \! dx_1 ~x_1\:G_{a/p} (x_1)\:G_{b/\bar p}(x_2)
\:\frac{4p_T}{2x_1-\overline{x}_T\:e^y}\:\frac{d\hat{\sg}}{d\hat{t}}
(ab\ra \ccbar[^{2S+1}L_J^{[1,8]}]\;d).
\eea
The contributing subprocess cross sections are
\bea
q ~\bar q &\ra& \qqbar[\slj] ~g, \non\\
g~ q (\bar q)  &\ra& \qqbar[\slj] ~q (\bar q),\\ 
g ~g &\ra& \qqbar[\slj] ~g,\non
\eea
where the $\qqbar$ is in the $^1 S^{[1]}_0$, $^1 P^{[8]}_1$ and 
$^3 S^{[8]}_1$ states.
The summation is over the partons, $a$ and $b$. $G_{a/p}$ and 
$G_{b/\overline{p}}$ are the distributions of partons $a$ and $b$ in the
hadrons and $x_1$ and $x_2$ are the momentum they carry, respectively.
$x_2$ is given in terms of $x_1$ as
\bea
x_2=\frac{x_1\:\overline{x}_T\:e^{-y}-2\tau}{2x_1-\overline{x}_T\:e^{y}},
\eea
where \(\tau=M^2/s\), with $M$ the mass of the resonance, $\sqrt{s}$ the
centre-of-mass energy, $y$ the rapidity at which the resonance is
produced and $\overline{x}_T=\sqrt{x_T^2+4\tau} \equiv 2 M_T/\sqrt{s}$, 
with $x_T=2p_T/\sqrt{s}$.

Among the coefficients the matrix elements for the subprocesses corresponding 
to $F_1[^1S_0]$ are  available in \cite{f1} and those 
corresponding to $F_8[^3S_1]$ 
are calculated by Cho and Leibovich \cite{cho2}. We have calculated the
remaining coefficient $F_8[^1P_1]$. The packages FORM and MATHEMATICA are
used to calculate the matrix element squares of the subprocesses.

One important component at large-$p_T$ is the gluon fragmentation
to charmonium, as has been shown in the case of large-$p_T$ $J/\psi$
production \cite{frag}. It has been noted in Ref.~\cite{cho1} that
even in the fixed-order calculation of the $J/\psi$ cross
section, the diagrams that contribute at large-$p_T$ are fragmentation-like
diagrams in which a single gluon attaches itself to a $c \bar c$ pair. 
In our case, we find that while the $F_8[^3S_1]$ has a fragmentation-like
contribution at large-$p_T$ there is no such contribution for the 
$F_8[^1P_1]$.  Following Ref.~\cite{cho1}, we have also corrected the
cross section by the ratio of $Q^2$ evolved to the unevolved fragmentation
function to get the correct shape of the $p_T$ distribution at large $p_T$.

Heavy quark spin-symmetry is made use of in obtaining 
$\left< {\cal O}_n^{\eta_c}\right>$'s from the experimentally available
$\langle {\cal O}_n^{J/\psi}\rangle$'s.  Using this symmetry we get the 
following relations among $\langle {\cal O}_n^H \rangle$'s:
\bea
\left< 0 \right|{\cal O}_1^{\eta_c}[^1S_0]\left| 0\right>&=&
\left< 0 \right|{\cal O}_1^{J/\psi}[^3S_1]\left| 0 \right>
\;(1+ O(v^2)), \non\\
\left< 0 \right|{\cal O}_8^{\eta_c}[^1P_1]\left| 0\right>&=&
\left< 0 \right|{\cal O}_8^{J/\psi}[^3P_0]\left| 0 \right>
\;(1+ O(v^2)), \non\\
\left< 0 \right|{\cal O}_8^{\eta_c}[^3S_1]\left| 0\right>&=&
\left< 0 \right|{\cal O}_8^{J/\psi}[^1S_0]\left| 0 \right>
\;(1+ O(v^2)).
\label{Ovalues}
\eea
Hence, for the singlet matrix elements we have $\left< 0 
\right|{\cal O}_1^{J/\psi}[^3S_1]\left| 0 \right>=1.2$ GeV$^3$ 
\cite{cho1} and for the octet matrix elements extracted from
the CDF $J/\psi$ data we have $A_1+A_2 \equiv$
\(\frac{\left< 0 \right|{\cal O}_8^{J/\psi}[^3P_0]\left| 0 \right>}
{M_c^2}+
\frac{\left< 0 \right|{\cal O}_8^{J/\psi}[^1S_0]\left| 0 \right>}{3}
=(2.2\pm0.5)\times 10^{-2}\) GeV$^3$ \cite{cho2}. The CDF $J/\psi$
do not allow for a separate determination of the values of $A_1$ and
$A_2$ because the shapes of these two contributions to the $J/\psi$ $p_T$
distribution are almost identical. We use the fact that the sum $A_1+A_2$
is constrained by the CDF data, and for our
numerical predictions we assume that either $A_1$ or $A_2$
saturates
the sum. In one case, therefore, we have the maximum possible contribution 
from the $^3S^{[8]}_1$ channel and none from the $^1P^{[8]}_1$ channel, 
and in the other case the $^3S^{[8]}_1$ channel makes no contribution while 
the $^1P^{[8]}_1$ contributes its maximum. 

\begin{figure}[ht]
\vskip 0.5in\relax\noindent
          \relax{\includegraphics{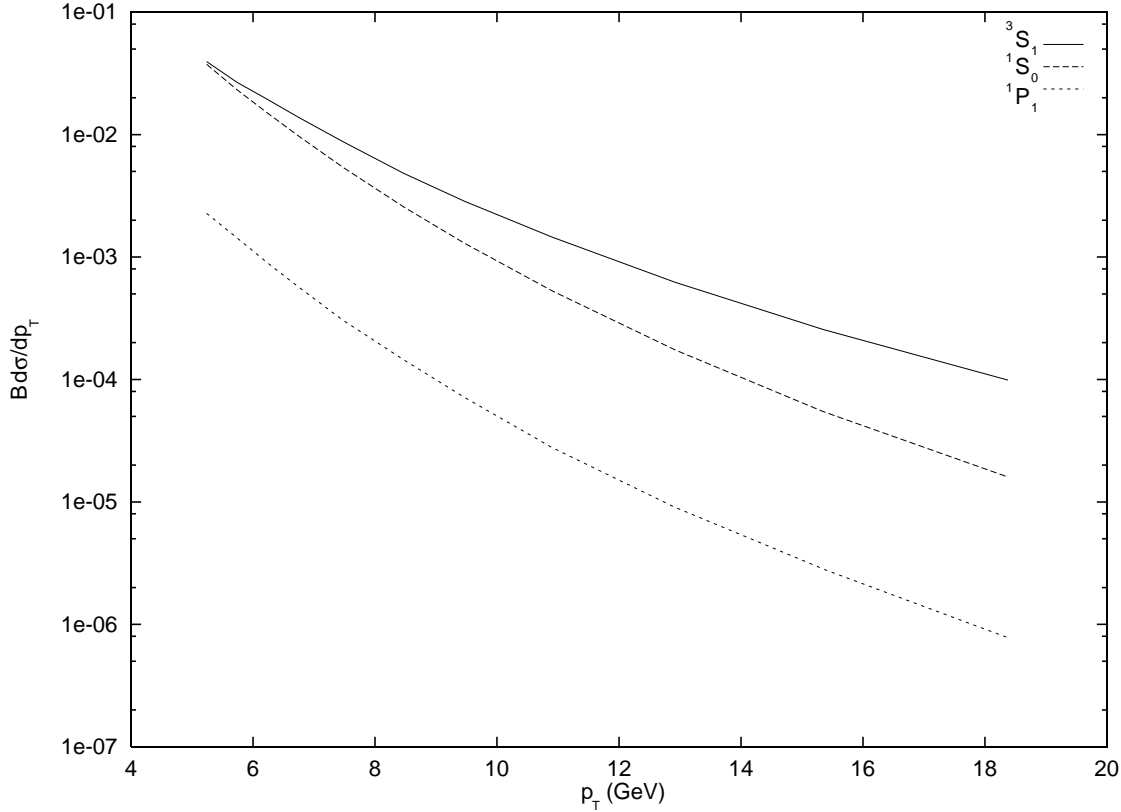}}  
\vskip 3.8in\relax\noindent
\caption{$d\sigma/dp_T$ (in nb/GeV) for $\eta_c$ production (after folding in
with Br($\eta_c \rightarrow \gamma\gamma)=3.0\times 10^{-4}$) in
$p \bar p$ collisions at 1.8~TeV with $-0.6 \le y \le 0.6$. }
\end{figure}

In Fig.~1, we have plotted the 
differential cross section $d\sg/dp_T$ against $p_T$. We are considering
the case where the $\eta_c$ will be searched for in its decay into
two photons and we have consequently folded in the differential
cross section with the $\eta_c \ra \gamma \gamma$ branching
ratio ($B=3 \times 10^{-4}$). In our computations, we have used 
MRSD-$^{\prime}$ parton densities \cite{mrs} and the parton densities 
are evolved to a scale $Q=M_T$. 
As a consequence
of the fact that we are saturating the sum with either $A_1$ or $A_2$,
our curves for the two octet channels represent the
maximum possible cross section that can result in each (with the
other being zero). 
Because of the fact that fragmentation dominates at high \pt\, 
contributions from $^1S_0^{[1]}$ and $^1P_1^{[8]}$, which
do not have a fragmentation contribution, fall steeply with increasing
\pt, whereas the $^3S_1^{[8]}$ contribution makes a sizable contribution
even at the largest $p_T$ considered. The shape of the measured $p_T$
distribution is clearly a very good measure of the individual matrix
elements $A_1$ and $A_2$. We reiterate that the CDF $J/\psi$ data
does not offer such a discrimination.

To get an idea of the measurability of the $\eta_c$ cross section
at the Tevatron, we have integrated our $p_T$ distribution for 
values of $p_T$ above 5~GeV. Assuming an integrated luminosity
of 110 pb$^{-1}$ we find that 
the number of events in the $\gamma \gamma$ channel from the singlet 
$^1S_0^{[1]}$ is about 400 and that from $^3S_1^{[8]}$ 
is about 7300 when $^1P_1^{[8]}$ contribution is absent, while the 
number of events
from $^1P_1^{[8]}$ is about 25 when $^3S_1^{[8]}$  contribution is absent. Thus
the number of $\eta_c \ra \gamma \gamma$ lies between 425 and 7700. Thus,
even the minimum of 425 events provides a very good prospect for the
observation of $\eta_c$ at the Tevatron. We also point out that the
integrated cross sections are also good measures of the matrix elements
$A_1$ and $A_2$. 

We have studied the effect on the cross section of the variation of
the parton densities, the scale and the non-perturbative matrix
elements. By using GRV HO densities \cite{grv}, instead of MRSD-$^{\prime}$,
we find that the cross section increases by about 25\%. If we use the
scale choice $Q=M_T/2$ instead of $Q=M_T$ then the cross section
increases by about 45-50\%. In addition to these uncertainties, we
can expect a 30\% variation in the values of the non-perturbative
matrix elements we have used, because of the fact that the heavy-quark 
symmetry is approximate.

In our calculations, we have considered only the direct production of 
$\eta_c$. There is a contribution to the $\eta_c$ signal coming
from the decays of $J/\psi$'s. From the measured $J/\psi$ cross
section at CDF we find that the contribution to the signal
coming from $J/\psi$ decays will be roughly about 100 events
which is only a small fraction of the minimum number of 
events that we expect from the direct production process.
A more complete analysis including the $J/\psi$ decay contribution
to the $\eta_c$ signal will be presented in a future publication.

We conclude with the following observations: the heavy-quark
symmetry of the NRQCD Lagrangian allows us to make predictions
for the $\eta_c$ cross section at the Tevatron. The integrated
rate and the shape of the $p_T$ distributions are very sensitive
to the non-perturbative matrix elements $A_1$ and $A_2$, whose
sum has been determined from the $J/\psi$ production data from
CDF, but are not known individually. This sensitivity can clearly
be used to determine these matrix elements individually. Our
estimates for the integrated rate vary between 425 and 7700
$\eta_c$ events (after including the branching ratio of the 
$\eta_c$ into two photons). Even at its minimum of 425 events,
the prospect of observability of the $\eta_c$ signal at the
Tevatron seems extremely good. We emphasise that this is a
testable prediction of NRQCD, and that such a prediction is
not possible in alternate approaches to quarkonium formation
like the colour evaporation model.

\end{document}